\documentclass[aps,prd,secnumarabic,amssymb, amsmath,nobibnotes,nofootinbib]{revtex4} %twocolumn
\usepackage{amsfonts,amsmath,hyperref,url, color}
\usepackage{bm, bbm}
\usepackage{graphicx}
\usepackage{mathtools}

\usepackage{t1enc}
\usepackage{hepnames}
\usepackage{adjustbox,booktabs}
\usepackage{makecell}

\newcommand{\bc}{\begin{center}}
\newcommand{\ec}{\end{center}}
\newcommand{\be}{\begin{eqnarray}}
\newcommand{\ee}{\end{eqnarray}}
\newcommand{\bs}{\begin{slide}}
\newcommand{\es}{\end{slide}}

\newcommand{\bi}{\begin{itemize}}
\newcommand{\ei}{\end{itemize}}

\begin{document}
\title{Planck-scale deformation of CPT and particle lifetimes}

\author{Michele Arzano}
\email{michele.arzano@na.infn.it}
\affiliation{Dipartimento di Fisica ``E. Pancini" and INFN, Universit\`a di Napoli Federico II, Via Cinthia,
80126 Fuorigrotta, Napoli, Italy}

\author{Jerzy Kowalski-Glikman}
\email{jerzy.kowalski-glikman@ift.uni.wroc.pl}
\affiliation{Institute for Theoretical Physics, University of Wroc\l{}aw, pl.\ M.\ Borna 9, 50-204
Wroc\l{}aw, Poland}
\affiliation{National Centre for Nuclear Research, ul. Pasteura 7, 02-093 Warsaw, Poland}

\author{Wojciech Wi\'{s}licki}
\email{wojciech.wislicki@ncbj.gov.pl}
\affiliation{National Centre for Nuclear Research, ul. Pasteura 7, 02-093 Warsaw, Poland}

%\date{\today}

\begin{abstract}
We carry out a systematic study of the bounds that can be set on Planck-scale deformations of relativistic symmetries and CPT from precision measurements of particle and antiparticle lifetimes. Elaborating on our earlier work \cite{Arzano:2019toz} we discuss a new form of departure from CPT invariance linked to the possibility of a non-trivial geometry of four-momentum and its consequences for the particle and antiparticle mass-shells and decay probabilities. Our main result is a collection of experimental bounds that can be obtained for the deformation parameter of the theoretical model under consideration based on current data and sensitivities of planned experiments at high energies.

\end{abstract}

\maketitle

\section{Theoretical considerations}

Discrete symmetries played a pivotal role in the theoretical and experimental development of particle physics. The CPT theorem, which ensures invariance of all physical phenomena under the combined action of the three discrete symmetries, is a direct consequence of the fundamental pillars of quantum field theory: locality and Lorentz invariance \cite{Streater:1989vi}. Any experimental departure from CPT invariance would contrast with such basic principles. As it stands, however, no experimental signatures contradicting invariance under CPT have been found \cite{Kostelecky:2008ts, Babusci:2013gda}.

A commonly accepted view is that at the Planck scale, i.e.\ at energies orders of magnitude higher than those accessible by current experiments, the quantum effects of gravity should drastically alter the structure of space-time in a regime in which the fate of locality and Lorentz invariance is unclear and thus that of CPT invariance.

In this work we continue our exploration \cite{Arzano:2019toz, Arzano:2016egk} of the effects of departures from CPT emerging in the presence of Planck-scale {\it deformations} of the algebra of relativistic symmetries, specifically we focus on the so-called $\kappa$-Poincar\'e algebra \cite{Lukierski:1991pn,Lukierski:1992dt,Lukierski:1993wxa, Majid:1994cy}.
This model has been studied in the past years as possibly encoding the physics of some {\it flat space-time limit} of quantum gravity. Its main feature is a four-momentum space which is no longer a vector space but a non-abelian Lie group. Such group is named $AN(3)$ (since three of its generators are abelian and one nilpotent) and can be described in terms of the Iwasawa decomposition of the five-dimensional Lorentz group $SO(4,1)\simeq SO(3,1)\, AN(3)$ \cite{KowalskiGlikman:2002ft, KowalskiGlikman:2003we}. This feature ensures a well-defined action of the four-dimensional Lorentz group $SO(3,1)$ on the group-valued momenta which geometrically are described by  points on a submanifold of the four-dimensional de Sitter space $dS_4$ defined in an embedding five-dimensional Minkowski space as
\be
-p_0^2 + p_1^2 + p_2^2 + p_3^2 + p_4^2 =\kappa^2\, ;\quad p_0+p_4>0\,.
\ee
The ``deformation parameter'' $\kappa$ is set by the curvature of the four-momentum manifold. In previous works \cite{Arzano:2019toz, Arzano:2016egk} we showed how the profound alterations that the new structure of four-momentum space brings in the context of quantum field theory lead to peculiar departures from CPT invariance which can be tested by looking at the lifetimes of particles and their antiparticles.
As it turns out, there are alternative ways to approach the question of deformation of CPT symmetry in the $\kappa$-deformed context. One can use the algebraic way, starting with the commutator algebra of the $\cal C$, $\cal P$, and $\cal T$ operators with Poincar\'e generators. This approach was adopted in the paper \cite{Arzano:2016egk}, where this commutator algebra was deformed with the help of the non-trivial Hopf algebraic concept of the antipode $S$ being the generalized minus and co-product, defining the action of the classical Poncar\'e algebra generators on tensor products of states.

Alternatively, one can take as a starting point the deformed, free quantum field. In what follows we will choose this route here, borrowing the results from the upcoming paper \cite{Arzano:2020xxx}. In this work we constructed the $\kappa$-deformed theory of a free complex scalar field. The result that concerns us here in the context of CPT symmetry is that the mass-shell relations for particles and antiparticles are not identical, but deformed in a subtle way. Let us describe this in details.

When embedding coordinates on the momentum manifold are taken as four-momenta of a particle, the associated translation generators $P_{\mu}$ are elements of the $\kappa$-Poincar\'e algebra expressed in the ``classical basis"\cite{KowalskiGlikman:2002we, Borowiec:2009vb} since the action of the Lorentz generators on $P_{\mu}$ on one-particle states labelled by four-momenta $p_{\mu}$ is the standard one and the mass Casimir invariant of the algebra is undeformed: $P_0^2-{\mathbf{P}}^2= m^2$.  A complete characterization of the space-time symmetries of the theory, however, requires a specification of their action on tensor products of one-particle states, describing systems with many particles. In the standard Poincar\'e case the action of symmetry generators follows the Leibniz rule of derivatives resulting in  the total momentum of a two-particles state being the sum of the momenta carried by its constituents. In the $\kappa$-Poincar\'e context one finds instead that the addition rules for spatial momenta and energies are given by
\begin{align}\label{sumk}
 \left(p^{(1)}\oplus p^{(2)}\right)_i &=
 p_i^{(1)}\frac{E^{(2)}+p_4^{(2)}}\kappa+p_i^{(2)}\,, \\ \left(E^{(1)}\oplus E^{(2)}\right) &=\frac1\kappa\, E^{(1)}\left(E^{(2)}+p_4^{(2)}\right) + \frac{\kappa E^{(2)}+\mathbf{p}^{(1)}\cdot \mathbf{p}^{(2)}}{E^{(1)}+p_4^{(1)}},
\end{align}
where
\begin{equation}\label{p4}
  p_4 = \sqrt{E^2-\mathbf{p}^2 +\kappa^2}\,.
\end{equation}
 This non-trivial composition of momenta is a direct consequence of the non-Abelian group law of $AN(3)$  (see ref. \cite{Kowalski-Glikman:2017ifs} for more details.)
Such $\kappa$-deformed rule of composition of four-momenta \eqref{sumk} is directly related to the noncommutative structure of the $\kappa$-Minkowski space-time \cite{Majid:1994cy} ``dual'' to the $AN(3)$ momentum space. The spacetime noncommutativity is encoded in the star product $\star$ of functions on spacetime which for plane waves takes the form
\begin{equation}\label{starproduct}
  e^{i(E^{(1)}t-\mathbf{p}^{(1)}\mathbf{x})}\star e^{i(E^{(2)}t-\mathbf{p}^{(2)}\mathbf{x})}= e^{i((E^{(1)}\oplus E^{(2)})t-(\mathbf{p}^{(1)}\oplus\mathbf{p}^{(2)})\mathbf{x})}\,,
\end{equation}
with $\oplus$ defined in Eq. \eqref{sumk}. Associated to the deformed composition rule we have a notion of the deformed inverse $\ominus$ reflecting the $AN(3)$ group inversion law, for which $(\ominus p)_i \oplus p_i=p_i \oplus(\ominus p)_i =0$ where
\begin{equation}\label{ominus}
  \ominus p _i = -p_i \frac{\kappa }{E+p_4} \equiv S(p)_i\,.
\end{equation}
Such deformed inverse \eqref{ominus} is known as {\it antipode} and denoted by $S(p)_i\equiv \ominus p _i$. Now, we get to our main point: as mentioned above, the four-momenta of a particle of mass $m$ are related by an ordinary mass-shell relation
\begin{equation}\label{t1}
  p^2 - m^2 = E^2 - \mathbf{p}^2 - m^2 = 0.
\end{equation}
For the corresponding antiparticle however, the mass-shell relation is given  in terms of the antipode $\bar S$
\begin{equation}\label{t2}
  \bar S(p)^2 -m^2 = \bar S(E)^2 - \bar{ \mathbf{S}}(\mathbf{p})^2 - m^2 = 0,
\end{equation}
where the components of the antipode are given by\footnote{This definition of $\bar S$  differs from the standard one of the antipode $S$ by a sign. This difference is a consequence of the fact that the momentum space of the antiparticle is a submanifold of de Sitter momentum space antipodal to the submanifold describing the momentum space of the particle. See Ref. \cite{Arzano:2020xxx} for details.} 
\begin{eqnarray}\label{t3}
 \bar S(E) & = & E - \frac{\mathbf{p}^2}{E +\sqrt{\kappa^2 +E^2 -\mathbf{p}^2}}, \nonumber \\
 \bar{ \mathbf{S}}(\mathbf{p}) & = & \frac{\kappa\mathbf{p}}{E +\sqrt{\kappa^2 +E^2 -\mathbf{p}^2}}.
\end{eqnarray}
One can check by the direct computation that the first equality in Eq. \eqref{t2} implies the second one, and vice versa.
This means that both the particle and the antiparticle belong to the same mass-shell manifold.
Moreover, when the particle is at rest, its momentum $\mathbf{p}$ vanishes and $E=m$.
When the antiparticle is at rest its momentum $\mathbf{\bar S}(\mathbf{p})$ again is equal to zero and $\bar S(E)=m$.

\section{Phenomenology of CPT deformation}

Consider now a particle and its antiparticle at rest in the laboratory frame.
For an observer boosted with respect to this frame with a boost parameter $\gamma$ the particle has energy and momentum
\begin{eqnarray}\label{t4}
  E & = & m\,\gamma, \nonumber \\
  |\mathbf{p}| & = & m\,\sqrt{\gamma^2 -1}.
\end{eqnarray}
For the antiparticle, however, the situation is different. According to Eq. \eqref{t3}, its energy and momentum are equal to
\begin{eqnarray}\label{t5}
 \bar  S(E) & = & m\,\gamma - \frac{m^2(\gamma^2 -1)}{m\,\gamma +\sqrt{\kappa^2 + m^2}}, \nonumber \\
  \mathbf{\bar S}(|\mathbf{p}|) & = & \frac{\kappa m\,\sqrt{\gamma^2 -1}}{m\,\gamma +\sqrt{\kappa^2 + m^2}}.
\end{eqnarray}
We can express the boost parameter $\gamma$  as a function of energy of the particle $\gamma = E/m$ obtaining for the energy of the antiparticle
\begin{eqnarray}\label{t6}
  \bar  S(E) & = & E - \frac{E^2(1 -1/\gamma^2)}{E +\sqrt{\kappa^2 + m^2}} \nonumber \\
             & \approx & E -\frac{E^2}{\kappa}\left(1-\frac1{\gamma^2}\right) \nonumber \\
             & = & E -\frac{\mathbf{p}^2}{\kappa}.
\end{eqnarray}

%\section{Phenomenological consequences}

%Similarly to the undeformed case, the deformed CPT operator $\Theta_\kappa=-S(p)$ is involutive, i.e. $\Theta_\kappa^{-1}=\Theta_\kappa$. Acting twice, $\Theta_\kappa$ reproduces the initial state's quantum numbers and four-momenta. However, even acting twice on the same state, its lifetime is always dilated after transforming it by $\Theta_\kappa$ or $\Theta_\kappa^2$
%Thus in a moving frame because deformation of CPT modifies its energy and thus the Lorentz boost
%\begin{eqnarray} \label{gamma}
%\gamma=\frac{E}{m} \xrightarrow{\Theta_\kappa} -\frac{S(E)}{m}= \frac{1}{m}(E-\mathbf{p}^2/\kappa).
%\end{eqnarray}
This subtle difference between the energy of the particle and its antiparticle \eqref{t6} reflects on different Lorentz factors affecting their lifetimes when passing from the rest frames to the laboratory frame. These Lorentz boosts are equal to $E/m$ for the particle and $\bar S(E)/m$ for antiparticle, leading to the different decay probabilities
\begin{eqnarray}\label{decay}
{\cal P}_{\mbox{\scriptsize part}} & = & \frac{\Gamma E}m\,\exp\left(-\Gamma t\, \frac Em\right), \nonumber \\
{\cal P}_{\mbox{\scriptsize apart}} & = & \Gamma\left(\frac Em - \frac{\mathbf p^2}{\kappa m}\right) e^{-\Gamma t \left(\frac Em - \frac{\mathbf p^2}{\kappa m}\right)},
\end{eqnarray}
where $t$ denotes the particle's proper time and $\Gamma=1/\tau$ stands for its decay width. At the phenomenological level this differences can be tested using precision measurements of lifetimes of particles and antiparticles in experiments at very high energies. Natural candidates for such measurements are the particle-antiparticle pairs, where both objects in a pair are their mutual CPT images, originating from two-body decays of resonances produced on existing or planned accelerators: the Large Hadron Collider (LHC) and the Future Circular Collider (FCC) \cite{fcc}, both at CERN.
Such experimental setting is advantageous for our study. Kinematics ensures that in the absence of deformation both particles have the same energies and fly back-to-back with equal momenta in the resonance rest frame. Transforming both particles to the laboratory frame and choosing pairs with the same transverse momenta one selects samples suitable for this study. Since an undeformed CPT transformation leaves energies and three-momenta intact, testing CPT invariance using pairs of CPT-coupled particles requires that their momenta are equal. Particles in pairs originating from resonance decays have non-zero transverse momenta such that even boosted using large Lorentz $\gamma$, their momenta remain slightly divergent with small opening angles $\theta$, of the order $10^{-6}-10^{-8}$ rad for energies considered here.
In order to make the pair strictly CPT-invariant and thus get rid of any differences besides those of the deformation, the momenta should be aligned by rotating them by $\pm\theta$. This is equivalent to assuming neither CPT- nor Lorentz violation due to anisotropy of space. The latter hypothesis is well supported by experimental results on search for a possible sidereal-time-dependence of interference of neutral $K^0$, $D^0$ or $B^0$ mesons, where no existence of an absolute direction in space has been detected with accuracy comparable or better than considered here, and with much larger rotations in space \cite{Babusci:2013gda,lhcb_cpt,babar_cpt,d0_cpt,focus_cpt}.

In Tab. \ref{Table:tab1} we present data on several possible decay channels, Lorentz boosts for energies attainable at LHC and FCC, experimental errors and limits on deformations $\kappa$. Crucial for this study is to have very high energy and not too heavy particles, since $1/\kappa$ corrections to lifetimes are proportional to $\mathbf p^2$ and to the reciprocal of mass. Concerning experimental accuracy, the critical quantity is an overall uncertainty of the particle lifetime $\tau$ that limits sensitivity of the measurement of the actual value of $\kappa$.

It is worthwile to provide a forecast of limits not only for the lightest particle-antiparticle pairs with the most accurately known lifetimes, like muons, but also for heavier decay products. As seen from Tab. \ref{Table:tab1}, limits on $\kappa$ obtained from the muons and pions are of the same order ($10^{14}$ GeV for LHC and $10^{16}$ GeV for FCC) whereas for tau leptons and mesons containing heavy quarks they are only an order of magnitude weaker. Lighter unstable parent resonances with masses around or below 1 GeV, like $\phi(1020)^0$, $\rho(770)^0$, $\omega(782)^0$ or $K_S$, can be more copiously produced in primary collisions and thus provide better statistical accuracy. Heavier resonances, like $D^0$, $B^0$, $\psi(3770)$ and heavy bottomonia $\Upsilon(10580)$ and $\Upsilon(10860)$, may decay into strange, charm or beauty mesons that live significantly shorter than muons or pions. As seen in Tab. \ref{Table:tab1}, for ultrarelativistic energies the $\gamma$ factor can be as large as $10^3-10^4$ at LHC and $10^4-10^5$ at FCC. This means that for a long-living particle, e.g. muon with $\tau=10^{-6}$ s, an accurate determination of particle's lifetime in laboratory requires an experimental baseline to be very long since $\gamma c\tau = 10^5-10^6$~m at LHC and $10^6-10^7$~m at FCC energies. For short-lived particles, like $B^\pm$ living $10^{-15}$ s, even large Lorentz boosts correspond to $\gamma c \tau$ not exceeding centimetres which could make the measurements easier.

\onecolumngrid

\vspace{.5cm}

\begin{table}[h]
	\caption{\em Limits on deformation parameter $\kappa$ for set of particle-antiparticle pairs and energies attainable at LHC and FCC. Values of decay times with errors and particle masses are from Ref. \cite{pdg}. In calculations of limits for $\kappa$ the lifetime accuracies were assumed everywhere $\frac{\sigma_\tau}{\tau}=10^{-6}$. Lorentz boosts $\gamma$ were calculated for energies 6.5 TeV for LHC and 50 TeV for FCC.}
	\label{Table:tab1}
	\vspace{5mm}
 \begin{adjustbox}{angle=0}
	\begin{tabular}{| c | c | c | c | c | c | c | c | c | c |}
		\hline
		\addlinespace[2mm]
		\scriptsize Particle & \makecell{\scriptsize Parent \\ \scriptsize resonance} & $\tau$[s] & $M$ [GeV] & $\frac{\Gamma}{M}$ & \makecell{$\frac{\sigma_\tau}{\tau}$ \\ \scriptsize (from PDG)} & \makecell{$\gamma$ \\ \scriptsize (LHC)}   & \makecell{$\gamma$ \\ \scriptsize (FCC)}  & \makecell{$\kappa=\frac{p^2}{m\delta_\tau}$ \\ \scriptsize(LHC)} & \makecell{$\kappa=\frac{p^2}{m\delta_\tau}$ \\ \scriptsize (FCC)} \\
		\addlinespace[2mm]
		\hline\hline
		
		\addlinespace[2mm]
		$\mu^\pm$ & $J/\psi, \Upsilon$ & $2.2\times 10^{-6}$ & $0.11$ & $2.8\times 10^{-18}$ & $1\times 10^{-6}$ & $6.1\times 10^4$ & $4.7\times 10^5$ & $4\times 10^{14}$ & $2\times 10^{16}$\\
		\addlinespace[2mm]
		\hline
		
		\addlinespace[2mm]
		$\tau^\pm$ & $J/\psi, \Upsilon$ & $2.9\times 10^{-13}$ & $1.8$ & $1.3\times 10^{-12}$ & $1.7\times 10^{-3}$ & $3.6\times 10^3$ & $2.8\times 10^4$ & $2.5\times 10^{13}$ & $1.5\times 10^{15}$\\
		\addlinespace[2mm]
		\hline
		
		\addlinespace[2mm]
		$\pi^\pm$ & \makecell{$K_S, \rho^0, \omega^0$ \\ $D^0, B^0$} & $2.6\times 10^{-8}$ & $0.14$ & $1.8\times 10^{-16}$ & $1.9\times 10^{-4}$ & $4.6\times 10^4$ & $3.6\times 10^5$ & $3\times 10^{14}$ & $1.8\times 10^{16}$\\
		\addlinespace[2mm]
		\hline
		
		\addlinespace[2mm]
		$K^\pm$ & $\phi^0, D^0, B^0$ & $1.2\times 10^{-8}$ & $0.49$ & $1.1\times 10^{-12}$ & $1.6\times 10^{-3}$ & $1.3\times 10^4$ & $1.0\times 10^5$ & $8.5\times 10^{13}$ & $5.1\times 10^{15}$\\
		\addlinespace[2mm]
		\hline
		
		\addlinespace[2mm]
		$D^\pm$ & $\psi, B^0$ & $1.0\times 10^{-12}$ & $1.9$ & $3.4\times 10^{-13}$ & $6.7\times 10^{-3}$ & $3.5\times 10^3$ & $2.7\times 10^4$ & $2.2\times 10^{13}$ & $1.3\times 10^{15}$\\
		\addlinespace[2mm]
		\hline
		
		\addlinespace[2mm]
		$B^\pm$ & $\Upsilon$ & $1.6\times 10^{-15}$ & $5.3$ & $0.8\times 10^{-13}$ & $2.4\times 10^{-3}$ & $1.2\times 10^3$ & $0.9\times 10^4$ & $0.8\times 10^{13}$ & $0.5\times 10^{15}$\\
		\addlinespace[2mm]
		\hline
		
	\end{tabular}
\end{adjustbox}
\end{table}

The best experimental accuracy of a mean lifetime, $10^{-6}$, is achieved for muons. The most significant contribution to this number comes from a dedicated measurement at low energy where despite the very large event sample the total error is dominated by the number of events: the statistical error is larger by a factor 3 than the systematic one \cite{tischchenko_2013}. For other particles discussed here: $\tau^\pm, \pi^\pm, K^\pm, D^\pm$ and $B^\pm$, present experimental accuracy is two or three orders of magnitude worse but, if needed, can be improved in currently working flavour factories like BELLE-II, BES-III and experiments at LHC. Therefore, in our estimates we assumed the relative lifetime accuracy of $10^{-6}$ for each particle. 

Currently used experimental techniques provide excellent time resolution. For instance, in the spectrometer used by LHCb it amounts to 45 fs for wide spectrum of momenta \cite{lhcb_detector} thus ensuring that systematic contribution to measurements of lifetimes of charged particles should be negligible compared to the statistical one.

A clear distinction between particle and antiparticle is possible due to their decays by using the charge of decay products. In most cases these are leptonic, semileptonic or hadronic decays where charge of the final-state lepton or hadron is equal to that of its parents, e.g. $\mu^\pm\rightarrow e^\pm\nu_e\nu_\mu$, $\pi^\pm\rightarrow\mu^\pm\nu_\mu$, $K^\pm\rightarrow l^\pm\nu_l$ ($l$ standing for $e$ or $\mu$), $K^\pm\rightarrow \pi^\pm X$, etc. If not all decays can be easily identified for experimental reasons, like worse identification or reconstruction efficiency for some particles, one has to take partial widths and modify decay times accordingly.
In Tab.~\ref{Table:tab1} we do not list all potentially interesting decay channels. We omit very rare ones, e.g. $J/\psi(2S)\rightarrow n\bar n, \Lambda^0\bar\Lambda^0, K^+K^-$, or those with neutral final states where the distinction between the particle and antiparticle would render it additionally demanding, e.g. if the final states do not contain CPT-coupled particles, as for $\phi^0\rightarrow K_L K_S$, or require more complex identification of neutral final-state particles via decay chains, as for example for $\psi(3770)\rightarrow D^0\bar D^0$, $\Upsilon\rightarrow B^0\bar B^0$ or $\Upsilon(10860)\rightarrow B_s\bar B_s$.

The eigenstates of CPT, such as $\pi^0$, represent a special case. Since $\pi^0$ is its own antiparticle, in a two-pion state created in the decay $K_S\rightarrow \pi^0\pi^0$, pions can be distinguished only kinematically, according to the four-momentum transformation $-S(p)$, but their quantum numbers remain the same.
Predictions for $\kappa$ from $K_S\rightarrow \pi^0\pi^0$ decay are close to that for the $K_S\rightarrow\pi^+\pi^-$, although for the neutral channel the lifetime is twice that big and its relative accuracy is twice worse, compared to the charged channel.

Let us now elaborate on another important point of our analysis. As noticed long time ago in Ref. \cite{khalfin}, decay laws have to account for non-zero widths of mass distributions by using the propagator of decaying particle. Thus the time- and momentum-dependent decay amplitude $a(t,\mathbf p)$ is given by a distribution $\omega(m; M,\Gamma)$ of the resonance mass $m$, with a mean mass $M$ and decay width $\Gamma$:
\begin{eqnarray}\label{a1}
a(t,\mathbf p)=\int_{-\infty}^\infty dm\,\omega(m; M,\Gamma)\,e^{-it\sqrt{m^2+\mathbf p^2}},
\end{eqnarray}
and the decay probability is equal to $\mathcal{P}(t,\mathbf p)=|a(t,\mathbf p)|^2$. In the non-relativistic approximation, the resonance mass distribution is approximated by the Breit-Wigner distribution
\begin{eqnarray} \label{a2}
\omega(m; M,\Gamma)= \frac{\Gamma}{2\pi}\frac{1}{(m-M)^2+(\Gamma/2)^2},
\end{eqnarray}
whereas in the more general, relativistic case it is given by
\begin{eqnarray} \label{a3}
\omega(m; M,\Gamma) & = & \frac{f(M,\Gamma)}{(m^2-M^2)^2+M^2\Gamma^2},
\end{eqnarray}
where $f(M,\Gamma)=\frac{2\sqrt{2}}{\pi}\frac{M\Gamma\sqrt{M^2(M^2+\Gamma^2)}}{\big[M^2+\sqrt{M^2(M^2+\Gamma^2)}\big]^{1/2}}$ and does not depend on $m$.
Some refinements of $\omega$ may be considered in the vicinity of thresholds for certain decay channels \cite{fonda_1978} but this could only negligibly affect our main findings here and we do not elaborate on them.
Since we are concerned here with ultrarelativistic particles, our discussion focuses on the relativistic Breit-Wigner distribution (\ref{a3}). Accounting for the non-trivial antipode $S(E)$ and using pole at $m^2=M^2+iM\Gamma$, one gets for $\omega$ given by Eq.(\ref{a3})
\begin{eqnarray} \label{a4}
a(t,\mathbf p) & = & \int dm\,\frac{f(M,\Gamma)}{(m^2-M^2)^2+M^2\Gamma^2} e^{-it(\sqrt{m^2+\mathbf p^2}-\mathbf p^2/\kappa)} \nonumber \\
	& = & e^{-it(\sqrt{M^2+\mathbf p^2+iM\Gamma}-\mathbf p^2/\kappa)}.
\end{eqnarray}
The new decay width $\tilde \Gamma$ can be calculated as imaginary part of the exponent's argument in Eq. (\ref{a4}).
Since the $\kappa$-dependent term contributes only to the real part of it, $\tilde\Gamma$ depends on $M$ and $\Gamma$ and momentum $\mathbf p$ but not on $\kappa$:
\begin{eqnarray} \label{a5}
\tilde\Gamma & = & 2\,\Im \Big(\sqrt{M^2+\mathbf p^2+iM\Gamma}-\frac{\mathbf p^2}{\kappa}\Big) \nonumber \\
& = & \sqrt{2} \Big[\sqrt{(M^2+\mathbf p^2)^2+M^2\Gamma^2}-(M^2+\Gamma^2) \Big]^{1/2}.
\end{eqnarray}
It has to be clarified here that independence of $\tilde\Gamma$ of $\kappa$ in Eq. (\ref{a5}) only means a lack of specific, deformation-dependent corrections to the decay width from using the Breit-Wigner distributions of the decaying particle mass instead of its sharp value. On the other hand, the correction term $\mathbf p^2/(2m\kappa)$ to $\Gamma$ in Eq. (\ref{decay}) is determined from the experimentally measured energy and momentum of the particle transformed using deformed CPT and it is always present in our model.

The formula (\ref{a5}) for the momentum- and mass-dependent decay width can be further simplified by using the smallness of the $\Gamma/M$ ratio, the latter amounting to the order $10^{-12}$ for $\tau^\pm$ and $10^{-16}$ for $\pi^\pm$ and $\mu^\pm$ (cf. Table \ref{Table:tab1}).
Expanding Eq.~(\ref{a5}) to linear terms in $\Gamma/M$ one gets
\begin{eqnarray} \label{a6}
\tilde\Gamma = \frac{M\Gamma}{\sqrt{M^2+\mathbf p^2}} + {\mathcal O}(\Gamma^2/M^2),
\end{eqnarray}
meaning that time in a moving frame is delayed by the Lorentz factor $\gamma=\sqrt{M^2+\mathbf p^2}/(M\Gamma)$.
This approximation is valid with good accuracy since $\Gamma/M$ is of the order $10^{-12}$ in the worst case of $\tau^\pm$ and $K^\pm,$ and $10^{-18}$ in the best case of $\mu^\pm$. It can be checked that terms proportional to $\Gamma^2/M^2$ scale like $1/\gamma^4$ thus giving corrections to $\tilde\Gamma$ not exceeding $10^{-12}-10^{-16}$ for LHC and $10^{-16}-10^{-20}$ for FCC. This means that {\it possible contributions to the decay time coming from non-exponential corrections to the decay law are by far smaller than present experimental accuracy} and thus we do not need to take them into account in our analysis.

In this paper we derived various bounds that can be set on the deformation parameter $\kappa$ describing Planck-scale relativistic kinematics as characterized by the $\kappa$-Poincar\'e algebra using precision measurements for lifetimes of particles vs. antiparticles. We found that the best bound from current data to be $\kappa  \lesssim (3-4)\times 10^{14}$ GeV obtained for measurements of lifetimes of muons and pions. Using future planned facilities this limit can be pushed up to $\kappa  \lesssim 2\times 10^{16}$ GeV. We pointed out that there is room for improvement in the sensitivity for measurements of other types of particles. Possible progress in detection techniques leading to better time resolution can move the limits on $\kappa$ even further. Particularly interesting would be imaging technologies based on femtosecond lasers that could hopefully improve time resolutions to be $\sim 1$ fs \cite{chen_2017}.

\section*{Acknowledgment}
The work of WW is supported by the Polish National Science Centre project number 2017/26/M/ST2/00697.  For JKG, this work was supported by funds provided by the Polish National Science Center, project numbers 2017/27/B/ST2/01902 and 2019/33/B/ST2/00050.

%For the non-relativistic Breit-Wigner distribution the formula for $\tilde\Gamma$, obtained for the first time in Ref. \cite{giacosa}, can be derived using distribution (\ref{a2}) in Eq. (\ref{a4}).
%For sake of completness we provide also the non-relativistic result for $\tilde\Gamma$, denoting it as $\tilde\Gamma_{\mbox{\scriptsize nrel}}$:
%\begin{eqnarray} \label{a7}
%\tilde\Gamma_{\mbox{\scriptsize nrel}} & = & 2\,\Im \Big(\sqrt{(M-i\Gamma/2)^2+\mathbf p^2}-\frac{\mathbf p^2}{\kappa}\Big) \nonumber \\
  %                                     & = & \sqrt{2} \Big[ \sqrt{(M^2+\mathbf p^2-\Gamma^2/4)^2+M^2\Gamma^2} \nonumber \\
   %                                    & - & M^2+\mathbf p^2-\Gamma^2/4\Big]^{1/2},
%\end{eqnarray}
%which, in approximation linear in $\Gamma/M$, is the same as Eq. (\ref{a6}). In this paper, however, all estimates were done using the relativistic formula (\ref{a5}).

%\vspace{2cm}

\end{document}